\documentclass[twocolumn,aps,prl,showpacs,superscriptaddress]{revtex4-1}
\usepackage{lmodern}

\usepackage[latin9]{inputenc} 
\usepackage{amsmath}
\usepackage{amssymb}
\usepackage{graphicx}
\usepackage{epstopdf}
\usepackage{enumerate}
\usepackage{ulem}
\usepackage{multirow}
\usepackage{xfrac}
\usepackage[usenames, dvipsnames]{color}

\usepackage{hyperref}
\hypersetup{colorlinks=true,linkcolor=black,citecolor=black}

\usepackage{xr}
\definecolor{mypink2}{RGB}{219, 48, 122}

\newcommand{\sk}{$S^{\text{\textonehalf}}_{1k}$}
\newcommand{\noise}{$S^{\text{\textonehalf}}$}

\newcommand{\vth}{$V_{\text{th}}$}
\newcommand{\asd}{[A/$\sqrt{\text{Hz}}$]}

\newcommand{\ino}{a:InO}
\newcommand{\s}{\space}

\begin{document}

\bibliographystyle{apsrev}

\title{Excessive Noise as a Test for Many-Body Localization}

\author{I. Tamir}
\affiliation{Department of Condensed Matter Physics, The Weizmann Institute of Science, Rehovot 76100, Israel.}
\email{Corresponding author; idan.tamir@weizmann.ac.il}
\affiliation{Fachbereich Physik, Freie Universit\"{a}t Berlin, 14195 Berlin, Germany.}

\author{T. Levinson}
\affiliation{Department of Condensed Matter Physics, The Weizmann Institute of Science, Rehovot 76100, Israel.}
\author{F. Gorniaczyk}
\affiliation{Department of Condensed Matter Physics, The Weizmann Institute of Science, Rehovot 76100, Israel.}
\author{A. Doron}
\affiliation{Department of Condensed Matter Physics, The Weizmann Institute of Science, Rehovot 76100, Israel.}
\author{J. Lieb}
\affiliation{Department of Condensed Matter Physics, The Weizmann Institute of Science, Rehovot 76100, Israel.}
\author{D. Shahar}
\affiliation{Department of Condensed Matter Physics, The Weizmann Institute of Science, Rehovot 76100, Israel.}
\affiliation{Department of Physics, Columbia University, New York, New York 10027, USA.}

\begin{abstract}
Recent experimental reports suggested the existence of a finite-temperature insulator in the vicinity of the superconductor-insulator transition. The rapid decay of conductivity over a narrow temperature range was theoretically linked to both a finite-temperature transition to a many-body-localized state, and to a charge-Berezinskii Kosterlitz Thouless transition.
Here we report of low-frequency noise measurements of such insulators to test for many body localization. We observed a huge enhancement of the low-temperatures noise when exceeding a threshold voltage for nonlinear conductivity and discuss our results in light of the theoretical models.
\end{abstract}

\maketitle

\begin{figure*}
	\includegraphics[width=\textwidth]{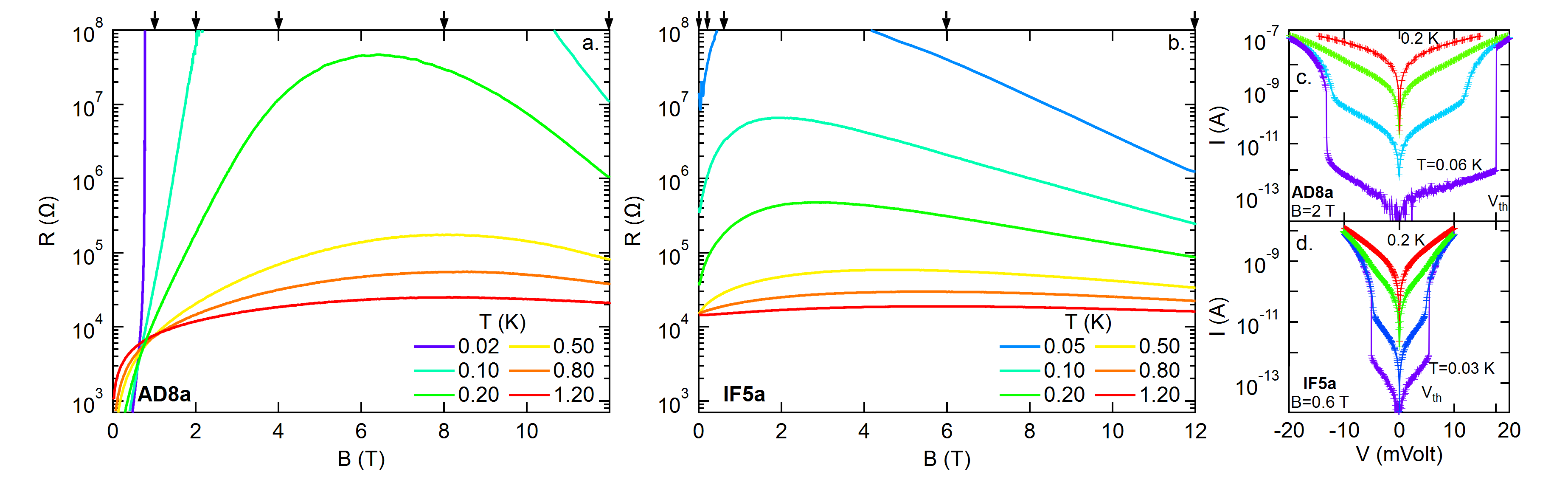}
    \caption{\textbf{Sample characterization.} a,b. $R$ vs. $B$ isotherms obtained from sample AD8a and IF5a respectively adopting a semi-log scale. The isotherms crossing in a. indicates the $B$-driven SIT in AD8a. An insulating peak is observed for both samples. The arrows indicate $B$ values used in Fig.~\ref{Vdep_diffBs}. c,d. $I$, in absolute value, vs. $V$ characteristic isotherms obtained from sample AD8a and IF5a respectively adopting a semi-log scale. The data in c. were measured at $B=2$ T and $T=0.06\space,0.1,\space0.15,\space0.2$ K. The data in d. were measured at $B=0.6$ T and $T=0.03\space,0.05,\space0.1,\space0.2$ K.} 
    \label{RB}
\end{figure*}

In certain thin-film superconductors, superconductivity is terminated by a transition to an insulating phase. This superconductor to insulator transition (SIT) can be experimentally driven by a variety of external parameters (for a review see Ref.~\cite{physupekhi}). In some cases, the SIT is followed by a strong insulating behavior, spanning a finite range of magnetic field ($B$), exhibiting an "insulating peak" \cite{paalanenprl69,GantmakherJETP96,murthyprl2004,vallesprl103,BaturinaPRL} (see Fig.~\ref{RB}a,b). Several theoretical works \cite{fisherduality,YonatanNat,PokrPrl,FeigAnnals,BouadimNat,Markus2013}, supported by experimental evidence \cite{GantmakherJETP98,murthyprl2004,STEINER200516,BaturinaPRL,craneprb751,benjaminprl101,vallesprl103,KopnovPRL,ShermanPRL,sacepe2015}, associate this peak with the localization of Cooper-pairs, terming this state a Cooper-pair insulator (CPI).

Close to the $B$-driven SIT, the insulating phase of amorphous indium oxide (\ino) thin-films exhibits faster-than-activated temperature ($T$) dependence of the sheet-resistance ($R$) as the conductivity approaches zero at a finite-$T$ \cite{ovadia2015}. This novel transition into a finite-$T$ insulating state gives way, at higher $B$'s, to sub-activated behavior consistent with the Efros-Shklovskii variable range hopping (VRH) mechanism of transport \cite{efros}. 

Qualitatively similar results were recently obtained from NbTiN thin-films \cite{mironov2018charge}. The authors of Ref.~\cite{mironov2018charge} relate the rapid decay in conductivity, at low $B$'s, to a charge-Berezinskii Kosterlitz Thouless (charge-BKT) transition \cite{PhysRevLett.65.645}, basing their analysis on the notion of vortex-charge duality \cite{fisherduality} that is predicted to govern the $B$-driven SIT in two-dimensional films. In this scenario, which is dual to the vortex-BKT transition in superconducting thin-films \cite{Berezinskii,KT,BKTinSC}, charge anti-charge unbinding above a critical $T$ ($T_{\text{c-BKT}}$) constitutes the main contribution of the measured conductivity thus explaining its rapid decay as $T$ approaches $T_{\text{c-BKT}}$.
It is interesting to note that in \ino\s thin-films, the faster-than-activated $R(T)$ insulating behavior severely violates duality-symmetry at low $T$'s \cite{MaozNat,PhysRevB.96.104513}.  

An alternative explanation for the precipitous drop of conductivity at a well-defined $T$ follows advancements in the field of many body localization (MBL). In 2005 it was shown \cite{mirlinprl,basko} that a system of isolated interacting electrons can undergo a finite-$T$ transition to a non-ergodic insulating MBL state. Basko, Aleiner and Altshuler (BAA) later proposed \cite{baskoprb} that this transition can also be observed in real, disordered, systems provided that the unavoidable electron-phonon (e-ph) coupling is sufficiently weak. In such a case, BAA argued, the MBL transition will manifest itself in nonlinear, bistable, current-voltage characteristics ($I-V$'s). Their theory further predicts a dramatic enhancement of the nonequilibrium current noise near the finite-$T$ transition.

Nonlinear, bistable, $I-V$'s were indeed observed in the low-$T$ ($T\lesssim0.3$ K) insulating phase of \ino\s thin-films \cite{murthyprl} (see Fig.~\ref{RB}c,d), as well as in other CPI's \cite{sanprb53,BaturinaPRL,mironov2018charge}.
These nonlinearities are associated \cite{maozprl} with an electron-overheating model that was introduced in Ref.~\cite{borisprl}. Within this model, under the application of external power ($P=V^2/R$) at low $T$'s, when the e-ph thermalization is ineffective, the electrons self-thermalize to a well defined $T$ ($T_{el}$) much greater than that of the host phonons. The nonlinear $I-V$'s are then related to changes in the strong $T_{el}$-dependence of $R$ which maintains an Ohmic relation: $R(T_{el})=V/I$. 
At even lower $T$'s, above some threshold $V$ (\vth), $I$ discontinuities result from bistable solutions of the heat balance equation:
$$\frac{V^2}{R(T_{el})}=\Gamma\Omega(T_{el}^\beta-T^\beta),$$
where $\Gamma$ is the e-ph coupling strength, $\Omega$ the sample volume, and the power $\beta\approx6$, calculated for metals in the dirty limit \cite{Schmid1974}, is found in experiment \cite{maozprl}. The success of this theoretical description \cite{maozprl,LevinsonPRB2016} provides an essential indication that at low $T$'s the electrons in our disordered \ino\s thin-films are decoupled from the phonons. 

To further investigate the feasibility of aforementioned theoretical possibilities we conducted a low-frequency ($f\le1$~kHz) noise study. Here we report the results of our extensive low-$T$ current power-spectral-density ($S$) measurements performed in the insulating phase of \ino\s thin-films as a function of different control parameters. Our main observation is a huge enhancement of $S$ when driving the system out of equilibrium and above \vth.

The results presented in this Letter are mainly obtained from two \ino\s thin-films, IF5a and AD8a \cite{Noise1}. IF5a is insulating at $B=0$ while AD8a is driven into an insulating phase by increasing $B$ above a critical value, $B_C=0.6$ T (see Fig.~\ref{RB}a,b where we plot $R$ vs $B$ isotherms obtained from both samples). The insulating peak, indicative of CPI's \cite{sacepe2015}, is observed in both samples.

The noise measured at relatively high $T$'s ($T\ge0.3$ K) can be fully explained considering noise sources generated by our experimental setup (for further discussion see Supplementary information). The total signal measured in this case can be described using the following form:
\begin{equation}
S_0=\frac{4k_BT}{dV/dI}+i_n^2+\bigg(\frac{v_n}{R_{\text{eff}}}\bigg)^2+\frac{k^2}{f}\enspace\bigg[\frac{\text{A}^2}{\text{Hz}}\bigg],
\label{S0}
\end{equation}
where $dV/dI$ is the sample differential resistance and $R_{\text{eff}}$ is the real part of the sample and measurement-wires parallel-impedance combination. At low-$f$'s $R_{\text{eff}}\approx dV/dI$. The first term in Eq.~\ref{S0} is the thermal noise, intrinsic to any resistive sample \cite{PhysRev.32.97,PhysRev.32.110} and independent of $f$ ("white noise"). All other terms are related to our measurement setup. Note that different units are used: $i_n$ \asd, $v_n$ [V/$\sqrt{\text{Hz}}$] and $k$ [A] $\propto (dV/dI)^{-1}$. 

At lower $T$'s our results are dramatically different. Initially, at $V$'s below \vth, we are unable to detect any noise above the instrumental noise (described by Eq. \ref{S0}). Above \vth\s the picture discontinuously transforms as the measured noise exhibits two successive $1/f^\alpha$ signals separated by a narrow peak.
The amplitude of the first $1/f^\alpha$ signal (at $f\lesssim 30$ Hz) follows the sharp decrease in $dV/dI$ that is observed above \vth\s(in accordance with the last term of Eq.~\ref{S0}). 
The appearance of an orders of magnitude increase in $S$, over a narrow $f$-range, followed by the second $1/f^\alpha$ decay is however unexpected. This observation is the focus of this Letter.

\begin{figure}
	\includegraphics[width=0.423\textwidth]{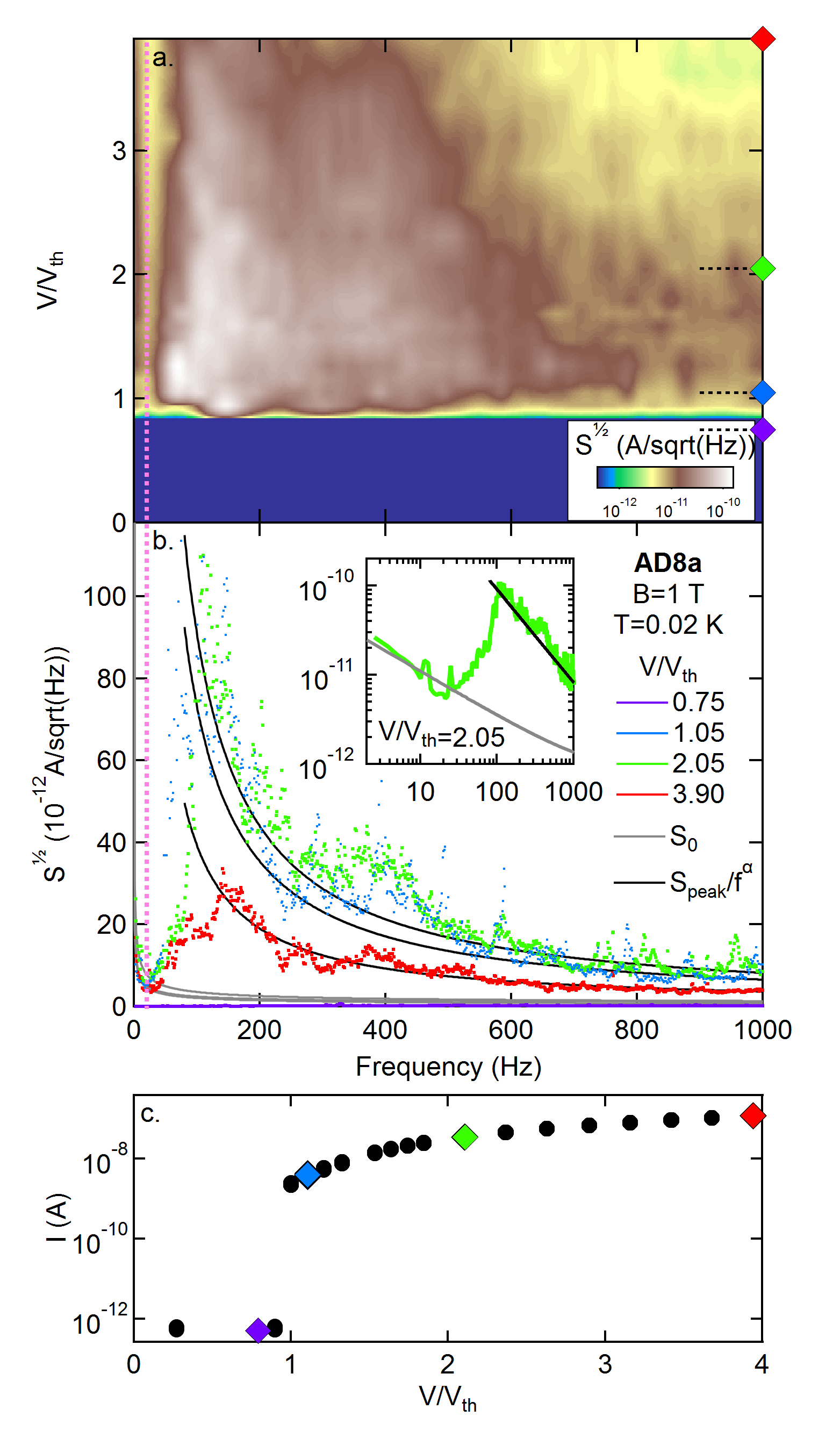}
    \caption{\textbf{Characteristic noise.} a. Color map of \noise\s vs. both $V$ (normalized by \vth\s \cite{Noise4}) and $f$, adopting a logarithmic color scale. b. Representative \noise\s vs. $f$ traces measured at different $V$'s (marked by diamonds in a,c). Fits to Eq.~\ref{S0} are plotted in gray. In both a,b. our data only follows Eq.~\ref{S0} over a narrow $f$-range (left of the vertical dashed pink line). Our best fits for the higher $f$'s data are plotted in black. Inset: Data measured at $V=2.05\cdot$\vth\s, adopting a log-log plot to stress the power-law nature of \noise\s decay. c. $I$ vs. $V$ measured simultaneously with the noise data adopting a semi-log plot. Data in a.-c. were obtained from the sample AD8a and measured at $B=1$ T and $T=0.02$ K.}
    \label{NoiseCM}
\end{figure}

Typical low-$T$ results are shown in Fig.~\ref{NoiseCM}a where we present a color map of the amplitude spectral density (\noise) vs. both $V$ and $f$. The data in the figure were obtained from AD8a at $B=1$ T and $T=0.02$ K. For reference we plot the, simultaneously measured, $I-V$ in Fig.~\ref{NoiseCM}c.    
To better illustrate the frequency dependence of the data we plot, in Fig.~\ref{NoiseCM}b, four \noise\s curves (dots) measured at several constant $V$'s. Our best fits to Eq.~\ref{S0} (solid gray lines) fail in describing the data. In fact, our data only follows Eq.~\ref{S0} over a narrow $f$-range (left to the dashed pink line). At higher $f$'s the data can be phenomenologically described by $S_{peak}/f^\alpha$ (solid black lines), where $S_{peak}$ is some fit parameter. To further stress the power-law dependence of \noise$(f)$ we re-plot, in the inset of Fig.~\ref{NoiseCM}b, the $V=2.05\cdot$\vth\s data adopting a log-log plot. Power-law dependences in such a plot appear as straight lines.

Before we continue our analysis, we note that $I$-discontinuities alone do not necessarily produce the measured noise spectra. In a control experiment that we conducted in the same experimental setup under similar conditions, but using a two-dimensional electron system residing in a Si-MOSFET that also exhibit discontinuous $I-V$'s, we did not detect any excessive noise. The results obtained form this experiment are presented in Fig.~S4 of the Supplementary information. 

\begin{figure*}
	\includegraphics[width=0.919\textwidth]{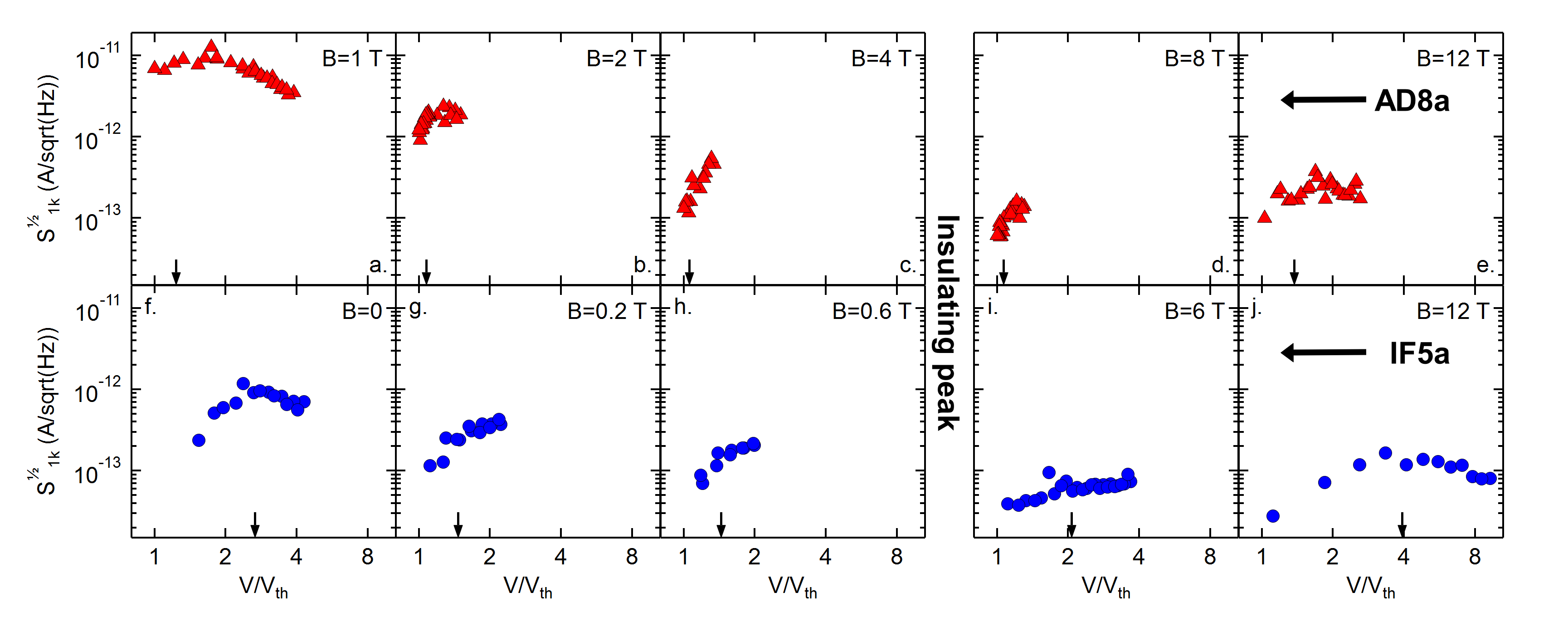}
	\caption{\textbf{Noise V dependence.} \sk\s vs $V$ (normalized by \vth\s \cite{Noise4}) obtained from AD8a (a.-e.), and IF5a (f.-j.), measured at  different $B$'s and at $T=0.02$ K, adopting a log-log plot. The arrows indicate $V/V_{\text{th}}$ values, used in Fig.~\ref{S1kAtp1j}, for which $j=0.01$ A/cm\textsuperscript{2}.}
	\label{Vdep_diffBs}
\end{figure*}

To facilitate a quantitative analysis of our results we define $S_{1k}$ to be the excessive noise evaluated at 1 kHz, $S_{1k}\equiv(S-S_0)|_{1\: \text{kHz}}$. While the peak amplitude of the noise would have been a more obvious choice, it exhibits non-trivial $f$ and time dependent fluctuations.
Using our definition of $S_{1k}$ we are able to described the measured data with high fidelity (see Supplementary information Fig.~S5). Several observations are now made possible. In Fig.~\ref{Vdep_diffBs} we present log-log plots of \sk\s {\it vs.} $V$ (normalized by \vth\s\cite{Noise4}) obtained from both samples at various $B$'s spanning both sides of the insulating peak (indicated by the arrows in Fig.~\ref{RB}a,b). Both samples exhibit qualitatively similar behavior: \sk\s is detected only above \vth, initially increasing with $V$. Further increase of $V$ is followed by a decrease in \sk, and, at even higher $V$'s, \sk\s drops below our measurement sensitivity (see Supplementary information Fig.~S5). 

\begin{figure}
	\includegraphics[width=0.423\textwidth]{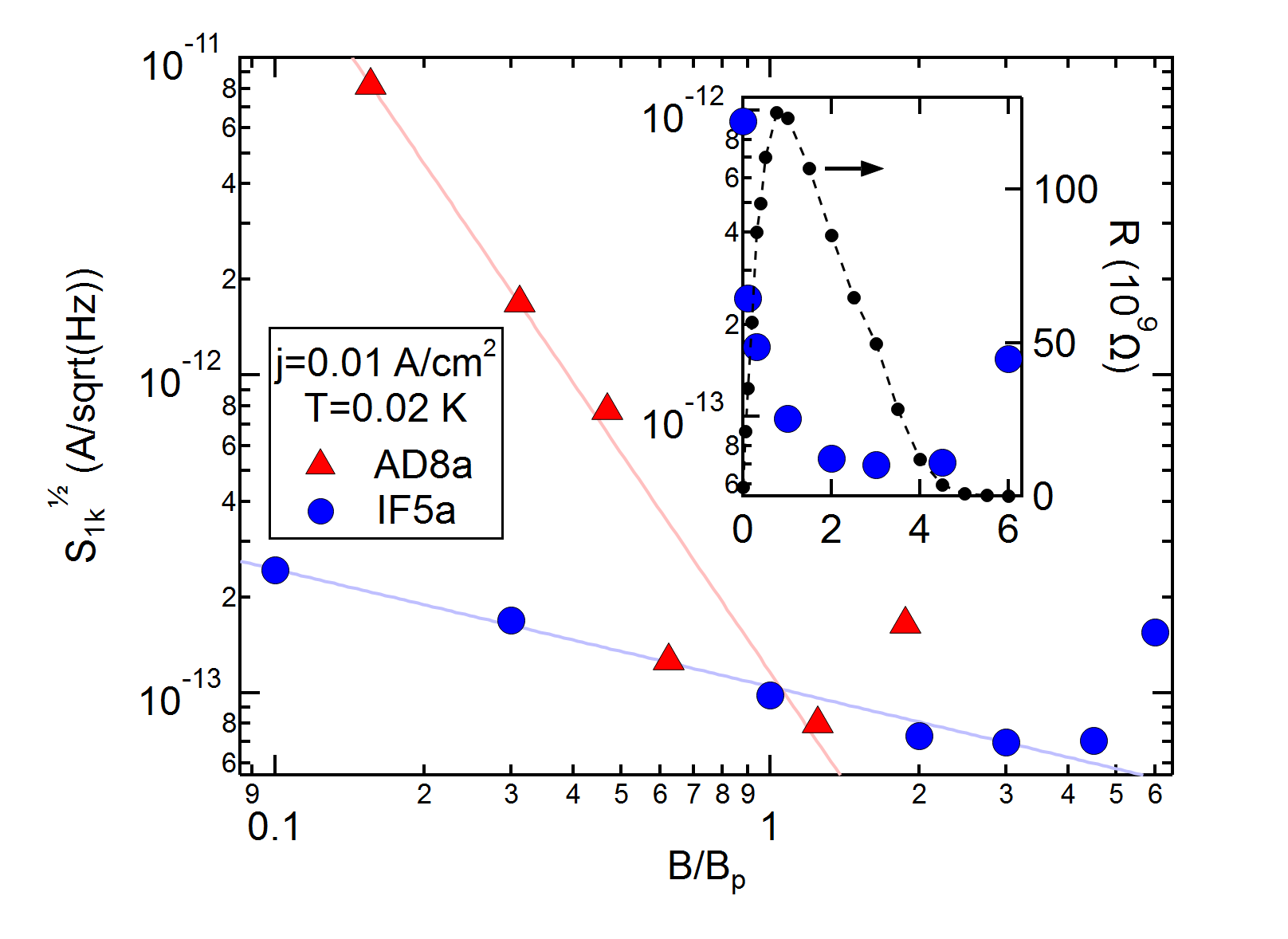}
	\caption{\textbf{Noise B dependence.} \sk\s, estimated at $j=$0.01 A/cm\textsuperscript{2}, vs. $B$ adopting a log-log plot. $B$ is normalized by $B_p=2,\;6$ T, the estimated value of the insulating peak for IF5a, AD8a respectively. Blue circles are used for IF5a and red triangles for AD8a. The lines are guides to the eye indicating the data decay at low to intermediate $B$'s. Inset: \sk\s vs. $B$ (blue circles, normalized by $B_p$) obtained from IF5a adopting a semi-log plot. $R$'s extrapolated from full $I-V$ scans \cite{ovadia2015} are plotted against the right axis (black circles). 
	}
	\label{S1kAtp1j}
\end{figure}

Next we address the $B$ dependence of \sk. Comparing the two samples, in Fig.~\ref{S1kAtp1j} we present a log-log plot of \sk\s vs. $B$, normalized by the estimated value of the insulating peak ($B_p$), evaluated at a constant $I$-density, $j=0.01$ A/cm\textsuperscript{2} \cite{Noise5}. We find that, for both samples, \sk\s decreases with $B$ (solid lines are guides for the eyes) for all but the highest $B$'s. It is worth noting that \sk\s does not follow the trend set by the insulating peak. This is demonstrated in the inset of Fig.~\ref{S1kAtp1j} where we plot $R$ {\it vs.}~$B/B_p$ (black circles, right axis), obtained from the sample IF5a at $T=0.02$ K, alongside \sk\s (blue circles, left axis) adopting a semi-log plot such that the $B=0$ data is also apparent.

We now wish to discuss our results in light of the theoretical models. Electron-overheating is unlikely the source of our observations since it is expected to manifest as a thermal, white, noise \cite{arai1983fundamental,PhysRevB.49.5942}. However, the data do not rule out electron-overheating in our system because the expected amplitude of this thermal noise component is below our sensitivity level.

The huge enhancement of $S$ above \vth\s supports a finite-$T$ transition to an insulating MBL state following BAA prediction \cite{baskoprb}. 
Furthermore, qualitatively similar spectral noise signature observed in superconducting films are related to vortex antivortex annihilation avalanches \cite{eggenhoffner2005vortex}. Analogously, our observation can indicate the occurrence of avalanche processes further supporting the possibility of a transition to an MBL state where avalanches are expected to occur near the finite-$T$ transition as many-electron cascades predominate the transport.
On the other hand, approaching the transition (by decreasing $V$), these cascades should involve more electrons and take longer times. Consequently, the peak in the power spectral density would shift to lower frequencies \cite{eggenhoffner2005vortex}. This reduction in peak $f$ is not supported by our measurements as is demonstrated in Fig.~\ref{NoiseCM}b. We are also unsure whether MBL can account for the excessive noise observed at high $B$'s where VRH dominates the transport. Finally, while we are unaware of any predictions suggesting noise enhancement as a result of the charge-BKT transition, relating our results to avalanche processes of charge anti-charge annihilations is tempting. A detailed understanding of this scenario awaits further theoretical developments.

In summary, we performed a set of low-$f$ noise measurements in the disorder- and $B$-driven insulating phases of \ino\s thin films. These measurements were devised as a test for a possible MBL transition previously suggested to govern the low-$T$ transport in such films. Above \vth\s we observed a sharp peak in $S$ which we attribute to the occurrence of avalanche processes that are dominating the conductivity. While this observation supports an MBL transition some discrepancies still prevent us from concluding the transition to such novel state, or excluding the possibility of a charge-BKT transition, without further theoretical and experimental considerations.

%
%

\section*{Acknowledgments}
We are grateful to I. Aleiner, B. Altshuler, M. Feigelman, A. Finkel'stein, and K. Michaeli for fruitful discussions. We are particularly grateful to D. Basko for valuable comments regarding MBL. This work was supported by the Israeli Science Foundation (ISF) Grant No. 751/13 and the Minerva Foundation, Federal German Ministry for Education and Research, Grant No. 712942. We also acknowledge the support provided by The Leona M. and Harry B. Helmsley Charitable Trust.

\bibliography{../../s1ahir}

\begin{thebibliography}{46}
\bibitem[{Noi({\natexlab{a}})}]{Noise1}
\bibinfo{note}{IF5a is 5 squares 0.1 mm wide bar and AD8a is 2 squares 2 mm
  wide bar. Both are 30 nm thick. Details regarding sample fabrication can be
  found elsewhere \cite{Shaharprb}.}
\bibitem[{Noi({\natexlab{b}})}]{Noise4}
\bibinfo{note}{Whenever the $I-V$'s are nonlinear but continuous we define
  \vth\s to be $V$ in which $dI/dV$ gains its maximum.}
\bibitem[{Noi({\natexlab{c}})}]{Noise5}
\bibinfo{note}{Similar $B$ dependence is also observed if, instead of
  evaluating \sk\s at a constant $j$, we do so at either a constant $I$, $V/I$,
  or $dV/dI$.}
\end{thebibliography}


\end{document}